# Compensated ferrimagnetic tetragonal Heusler thin films for antiferromagnetic spintronics

Roshnee Sahoo[1], Lukas Wollmann[1], Susanne Selle[2], Thomas Höche[2], Benedikt Ernst[1], Adel Kalache[1], Chandra Shekhar[1], Nitesh Kumar[1], Stanislav Chadov[1], Claudia Felser[1], Stuart S. P. Parkin[3], & Ajaya K. Nayak[1,3*]

[1]*Max Planck Institute for Chemical Physics of Solids, Nöthnitzer Str. 40, 01187 Dresden, Germany*
[2]*Fraunhofer Institute for Microstructure of Materials and Systems IMWS, Walter-Hülse-Str. 1, 06120 Halle, Germany*
[3]*Max Planck Institute of Microstructure Physics, Weinberg 2, 06120 Halle, Germany*

Spintronics is a large field of research that involves the generation, manipulation and detection of spin currents in magnetic heterostructures and the use of these currents to excite and to set the state of magnetic nano-elements.[1,2] The field of spintronics has focused on ferromagnetic thin film structures in which charge currents can be spin-polarized via interfacial and volume spin dependent scattering. However, ferromagnets produce magnetostatic dipole fields, which increase in size as devices are scaled to smaller dimensions. These must be minimized or eleminated to enable operation of spintronic field sensing and magnetic memories.[2] One way to do this is to take advantage of the phenomenon of long-range oscillatory interlayer coupling[3] to create synthetic antiferromagnetic heterostructures[1] or the use of ferrimagnetic materials such as rare-earth-transition metal alloys at their compensation point.[4] The latter ferrimagnetic materials are of special inerest because they can completely eliminate dipole fields volumetrically, even on the atomic scale. However, materials are needed that can operate over a wide temperature window and not solely at or in the vicinity of a compensation temperature. Here we show that Heusler compounds can be designed for this purpose.

Heusler compounds, $X_2YZ$ (where $X$, $Y$ are transition metals and $Z$ is a main-group element), are well known for their potential applications in spintronics, especially in spin-torque based devices.[5] These materials crystallize in both cubic and tetragonal crystal structures with multiple magnetic sub-lattices, and hence are good candidates for engineering a wide range of complex magnetic structures. In particular all the known tetragonal Heuslers are ferrimagnetic with at least two magnetic sub-lattices whose magnetizations are aligned anti-parallel to one another, or, as has been shown recently, can be non-collinear to one another. The Mn based tetragonal Heuslers are of particular interest because their magnetic ordering temperatures can be well above room temperature, but the magnetizations of their two sub-lattices are typically distinct, leading to a net uncompensated magnetization. One of the most interesting of these materials is $Mn_3Ga$ which has a Curie temperature of ~750 K.[6] By tuning the magnetization of the two sub-lattices in $Mn_3Ga$ by changes in composition, we propose that a fully compensated ferrimagnetic (CFI) Heusler is possible. To help identify the needed compositional variations we have carried out density functional calculations of the electronic structures of $Mn_{3-x}Y_xGa$ $(0 \leq x \leq 1)$ (at zero temperature) for the elements $Y =$ Ni, Cu, Rh, Pd, Ag, Ir, Pt, Au, which are non-magnetic or nearly non-magnetic when substituted into Mn3Ga. We show that in all these cases it is theoretically possible to obtain a CFI , which we have experimentally validated for the case of $Y$=Pt.

The present calculation is based on the experimental lattice parameters of the bulk $Mn_3Ga$ lattice.[6] We find that a compensated magnetic state can be achieved when the manganese atoms in the parent compound $Mn_3Ga$ are substituted by a late transition metal (LTM), $Y$, with a valence electron count (VEC) $N_V(Y) \geq 9$. As shown in Fig. 1, a compensation point (shown by the dotted line for Pt) can be found for $Y =$ Ni, Cu, Rh, Pd, Ag, Ir, Pt, Au, for a particular stoichiometry at $T$=0 K. In all theses cases the magnetic moment per formula unit decreases as Mn is substituted by *Y*. With the exception of Ag and Au, the moment monotonically decreases with increasing Y. For elements with the same number of valence electrons, the value of Y where the moment goes to zero is nearly the same. At this compensation point, even though the net moment vanishes, it is of interest that a considerable spin-polarization can still be achieved. This is a consequence of the symmetrically inequivalent Mn sub-lattices that, thereby, leads to distinct spin-dependent density of states at the Fermi energy (see supplementary information).

Previously, it has been shown in bulk materials that Mn-Pt-Ga forms a complex structure with multiple magnetic phases.[7] As shown in Fig. 2a the host compound $Mn_3Ga$ has a layered structure, with, along the tetragonal axis, aternating planes formed from Mn-Ga and Mn-Mn.[6-8] The Mn moments are oriented perpendicular to the respective planes but the moments have different magnitudes in the two planes. The Mn moments in the Mn-Ga plane are calculated to display a moment of $\approx 3.1\ \mu_B/f.u.$ which couple antiferromagnetically to the Mn moments in the Mn-Mn plane, which have a larger total moment of $\approx 4.2\ \mu_B/f.u.$ (Fig. 2a). By partially replacing Mn in the Mn-Mn plane by Pt, the effective moment in this plane can be reduced to match that of the moment in the Mn-Ga plane (Fig. 2b), thereby giving rise to a fully compensated magnetic state.

The practical application of these materials depends on the successful growth of thin films with a fully compensated magnetization and a high magnetic ordering temperature. Here we study the substitution of Mn by Pt in $Mn_{3-x}Pt_xGa$ thin films, and report the growth of tetragonal thin films with a tunable magnetic anisotropy and a fully compensated magnetic state in $Mn_{2.35}Pt_{0.65}Ga$. We also show that layers formed from this compensated ferrimagnet can be combined with an uncompensated ferrimagnetic (FI) material to achieve a large exchange bias (EB).

Thin films of composition $Mn_{3-x}Pt_xGa$ ($x$ =0.2, 0.5, 0.6, 0.65, 0.7) with thicknesses ranging from 15 to 50 nm were grown by DC magnetron sputtering on $SrTiO_3$ (STO)(001) substrates at a temperature of 350 °C. The films were co-sputtered using three targets of $Mn_{50}Ga_{50}$, Mn, and Pt for which the power to each target could be varied independently. The films exhibit (002) and (004) reflections in θ-2θ X-ray diffraction (XRD) patterns, signifying the epitaxial growth of a tetragonal Heusler structure. A detailed description of the structural analysis is presented in Fig. S1 and Table S1 of the supplementary information.

The $Mn_{3-x}Pt_xGa$ films display a tetragonal crystal structure over the entire compositional range that we considered. Hence, an out of plane magnetic anisotropy can be expected. Indeed we find from perpendicular magnetization $M(H)$ versus out of plane field hysteresis loops that these films exhibit a large out of plane magnetic anisotropy with large coercive fields, $H_C$. $M(H)$ loops measured at 50 K and 300 K for various thin films, are shown in Fig. 2c. The sample with a small Pt content, i.e, $x=0.2$, exhibits large coercive fields of 1.5 T at 50 K and 1.2 T at 300 K. Note that this sample has a weak dependence of the saturation magnetization with temperature (0.8 $\mu_B$/f.u. and 0.7 $\mu_B$/f.u. at 50 K and 300 K, respectively), indictaing a Curie temperature ($T_C$) that is well above room temperature. The $M(T)$ curves measured up to 400 K for 50 nm thick films shows that the magnetic ordering temperature is well above 400 K for all these samples. With increasing Pt concentration the magnetization decreases and becomes nearly zero for $x=0.65$. This fully compensated film with composition $Mn_{2.35}Pt_{0.65}Ga$ has a magnetization close to zero over the the complete temperature range of our measurements. The presence of a small anomaly in the $M(H)$ loops near zero field for many of our samples likely originates from a small in-plane magnetization component. The same feature has been very often observed in $Mn_{3-x}Ga$ thin films.[9,10]

We have demonstrated that the magnetic anisotropy in the tetragonal $Mn_{3-x}Pt_xGa$ thin films can be successfully tuned and a compensated magnetic state can be achieved. Now we focus on utilizing these compensated thin films for spintronic applications. It is well known that exchange bias (EB) is one of the most important tools that has been employed to spin-engineer a variety of useful spintronic devices.[1,11,12] To examine whether the present CFI films can be implemented in EB devices, we have designed a bilayer formed from a layer (Layer-1) of the compensated ferrimagnet, $Mn_{2.35}Pt_{0.65}Ga$, together with a layer of an uncompensated ferrimagnet layer (Layer –2) $Mn_{2.8}Pt_{0.2}Ga$, as schematically shown in Fig. 3a. The thickness of the CFI film is chosen to be 27 nm so as to lower the $T_C$ below 400 K (Supplementary Fig. S3b). Cross-sectional scanning transmission electrron microscopy (STEM) image of the bilayer film ( Fig. 3b) clearly displays two distinct layers formed from the FI and CFI Mn-Pt-Ga films. As shown in Fig. 3c, the high resolution TEM images taken from the bilayer film indicate epitaxial growth of the FI and CFI layers with the tetragonal axis perpendicular to the plane of the film. The diffraction patterns of Layer-1 and Layer-2 clearly demonstrate the tetragonal structure of both the CFI and FI layers (Fig. 3d). The absence of some minor diffraction spots in the FI film (Layer-2) might originate from reduced atomic order in $Mn_{2.8}Pt_{0.2}Ga$, as evident from the XRD measurement of the single layer films (Supplement Fig. S1a).

To characterize the magnetic properties of the bilayer films, we have measured zero field cooled (ZFC) and field cooled (FC) $M(H)$ loops at $T=5$ K, as shown in Fig. 3e. As can be seen, the ZFC $M(H)$ loop exhibits a symmetric behavior with respect to positive and negative fields so that the loop is centered near $H=0$. By contrast, the FC $M(H)$ loop is shifted considerably in the negative field direction when cooled with a field of +5 T from 400 K. When the field cooling procedure is repeated with a cooling field of -5 T, the $M(H)$ loop shifts in the opposite positive field direction by the same amount. Hence, it is clear that the CFI/FI bilayer displays a large exchange bias due to an exchange interaction between the CFI and FI layers. It is also important to note that exchange biasing the FI layer with the CFI layer slightly changes the magnetic state of the bilayer film, as evident from the enhanced magnetization of the FC $M(H)$ loop in comparision to that of the ZFC one.

The temperature dependence of the exchange bias field is determined from a series of $M(H)$ loops that are realized by field cooling in 5T to each temperature. The exchange bias field ($H_{EB}$) increases as the temperature is reduced and reaches 0.22 T at 5 K, as shown in Fig. 3f. $H_{EB}$ decreases monotonically with increasing temperature: a finite value of $H_{EB}$ is obtained at 300 K. We find that $H_{EB}$ increases as the magnitude of the FC field $H_{FC}$ is increased and saturates for $H_{FC}$ greater than 2 T (inset of Fig. 3f). Similar to $H_{EB}$, the coercive field $H_C$ also decreases with increasing temperature. It is important to mention that the bulk Mn-Pt-Ga system (when inhomogeneoous) shows a non-zero EB for temperatures up to only 150 K.[7]

We now focus on transport measurements to characterize the CFI state and EB in Mn-Pt-Ga thin films. It is well known that ferromagnetic (FM) materials exhibit an additional component to the normal Hall effect, namely an anomalous Hall effect (AHE). The AHE generally is absent for materials with no net magnetic moment. Therefore, Hall effect measuremens are an important means to characterize the present thin films. The Hall resistivity ($\rho_{xy}^A$) measured for several Mn-Pt-Ga thin films are shown in Fig. 4a. $Mn_{2.3}Pt_{0.7}Ga$, with a Pt concentration just below the compensation point, displays $\rho_{xy}^A \approx 0.16\ \mu\Omega$ cm (at zero field) at 2 K. For the magnetically compensated film $Mn_{2.35}Pt_{0.65}Ga$, $\rho_{xy}^A$ at zero field (the spontaneous component related to the magnetization) becomes nearly zero. By further decreasing the Pt concentration to $x=0.2$, $\rho_{xy}^A$ increases considerably, signifying a scaling of the AHE with the net magnetization of the Mn-Pt-Ga alloy.[13,14] It is interesting to note that $\rho_{xy}^A$ changes its sign from positive to negative as the composition is tuned through the compensation point. The substitution of Mn with Pt leads to a decrease of the net down-spin moment as the Pt atoms preferably occupy the Mn in Mn-Mn planes. By crossing through the compensation point the net up-spin moment dominates the magnetization state for $x = 0.7$, which can account for the change in sign of $\rho_{xy}^{A}$. This change in sign in the Hall effect measurement clearly indicates that the present system passes through a compensation point, when Pt is substituted for Mn in $Mn_{3-x}Pt_{x}Ga$.

The increase in the magnetization of the FC $M(H)$ loop in the bilayer film indicates that the exchange interaction at the interface also affects the total magnetic state of the sample. For more insight into this phenomenon we have measured the AHE of the bilayer film in ZFC and FC modes, as shown in Fig. 4b. In ZFC mode the AHE retraces the symmetric nature of the ZFC AHE of the single layer film. Since the AHE directly scales with the magnetization, the increase in magnetization in the FC mode is also visualized by a substantial increase of the AHE for otherwise similar measurement conditions. In addition, the FC AHE versus field hysteresis loop also exhibits a shift towards the negative field direction, just like the FC magnetization loop. Therefore, it is quite evident that there is indeed a change in the total magnetic state of the system when the ferrimagnetic layer is exchange coupled with the compensated layer. Note that the FC and ZFC experiments are repeated several times to exclude any thermal annealing or hysteretic effects.

The observation of an enhanced magnetization and AHE of the field cooled bilayer films suggests the appearance of an additional ferromagnetic component when the ferrimagnetic layer is exchange-biased with the compensated ferrimagnet. To understand the underlying phenomena we show a schematic diagram with the spin configurations for the different interfaces in Fig. 4c-e. Fig. 4c shows the idealized spin configuration of a regular FM/AFM EB system with FM interfacial exchange interaction. This type of exchange interaction has been explored in many FM/AFM EB systems.[15,16] In the present case the EB is obtained in a FI/CFI system, where both FI $Mn_{2.8}Pt_{0.2}Ga$ and CFI $Mn_{2.35}Pt_{0.65}Ga$ exhibit almost the same lattice parameter. Therefore, the formation of Mn-Pt (FI layer)/Mn-Ga (CFI layer) (Fig. 4d) or Mn-Ga (FI layer)/Mn-Pt (CFI layer) (Fig. 4e) interfaces are the most likely scenario. In this case, the system will behave as a single material at the interface with a continuous unit cell (as marked by the dashed box). However, the strong ferromagnetic exchange at the interface breaks the alternating plane to plane AFM ordering. All spins within each of the Mn-Ga planes should align in the same direction, but the exchange interaction between the moments in layer "spin-1" and layer "spin-3" should be FM in nature (the layers "spin-1" through "spin-5" are defined in Fig. 4d). On the other hand the strong AFM between moments in layer "spin-2" and layer "spin-3" opposes a FM interaction between layer s "spin-1" and "spin-3". As a result of these competing FM and AFM interactions the moments in the layer "spin-3" will reorient themselves to be aligned along an angle $\theta$ as shown by the dashed arrows in Fig. 4d. Since the next nearest exchange interaction becomes weaker with distance the angle $\theta$ also decreases for the moments in layer "spin-5" and, subsequently vanishes above a critical distance. A similar magnetic picture can be expected for the Mn-Ga and Mn-Pt interface, as shown in Fig. 4e. Due to canting of the spins in the FM layer, the total magnetization and anomalous Hall conductivity can increase in the field cooling process. This type of spin canting due to competing FM and AFM interactions has been recently observed in the $Mn_2RhSn$ Heusler compound.[17]

In recent years, antiferromagnetic spintronics has received much attention since ideal antiferromagnets do not produce stray fields and are much more stable to external magnetic fields compared to materials with net magnetization.[18-22] Akin to antiferromagnets, compensated ferrimagnets have zero net magnetization but have the potential for large spin-polarization and strong out of plane magnetic anisotropy, and, hence, are ideal candidates for high density memory applications. We have demonstrated that a fully compensated magnetic state with a tunable magnetic anisotropy can be realized in Mn-Pt-Ga based tetragonal Heusler thin films. Furthermore, we have shown that a bilayer formed from a fully compensated and a partially compensated Mn-Pt-Ga layer, exhibits a large interfacial exchange bias up to room temperature. The present work establishes a novel design principle for spintronic devices that are formed from materials with similar elemental compositions and nearly identical crystal and electronic structures. Such devices are of significant practical value due to their improved properties such as thermal stability and spin transport across the inteface.

Moreover, a single layer of the compensated tetragonal thin film could replace the synthetic antiferromagnet, which is widely used in field sensing spintronic devices[1] and which has been rececntly shown to give rise to highly efficient current driven motion of domain walls in the racetrack memory device.[23] The achievement of EB up to room temperature in the present bilayer films is an important step towards practical applications. Although exchange bias has been studied in some unconventional systems, such as, a ferromagnet/spin glass[24], a ferromagnet/paramagnet[25] and a natural mineral[26], the present exchange bias from ferrimagnet/compensated ferrimagnet is the first of its kind and has significant potential for technological applications. The flexible nature of Heusler materials to achieve tunable magnetizations, and anisotropies within closely matched materials provides a new direction to the growing field of antiferromagnetic spintronics.

## Experimental section

Thin films of Mn-Pt-Ga were grown by DC magnetron sputtering on MgO(001) at a substrate temperature of 350 °C using co-sputtered targets of $Mn_{50}Ga_{50}$, Mn, and Pt. The base pressure of the ultra-high vacuum (UHV) system was maintained below 1E-8 mbar. The films were subsequently annealed for 15 mins inside the sputtering system at 350 °C. Next, the temperature was reduced to room temperature, and a capping layer of 2-3 nm of Pt was deposited to prevent the film from oxidizing. The crystal structure and thickness of films was examined by means of X-ray diffraction (XRD) measurements using a Cu-$K_\alpha$ source (Philips PANanalytical X'pert Pro). We performed energy-dispersive X-ray spectroscopy (EDX) measurements to determine the composition of the samples. We also used the StrataGEM software package to obtain the film thickness. Transmission electron microscope (TEM) samples were prepared via focused ion beam (FIB) ex-situ lift-out. An athermal Pt-cover protected the sample from ion damage during FIB preparation. The FIB lamella was transferred ex-situ with a micromanipulator on a TEM carbon grid. TEM investigation was performed in a FEI TITAN G2 80-300 microscope at 300 keV. The magnetization measurements were carried out on a vibrating sample magnetometer (MPMS 3, Quantum Design). The transport measurements were performed with low-frequency alternating current (ACT option, PPMS 9, Quantum Design). The theoretical calculation of spin-polarization were performed by using the SPR-KKR method and by treating disorderin terms of CPA (coherent potential approximation).

**Acknowledgements:** This work was financially supported by the ERC Advanced Grant No. (291472) "Idea Heusler".

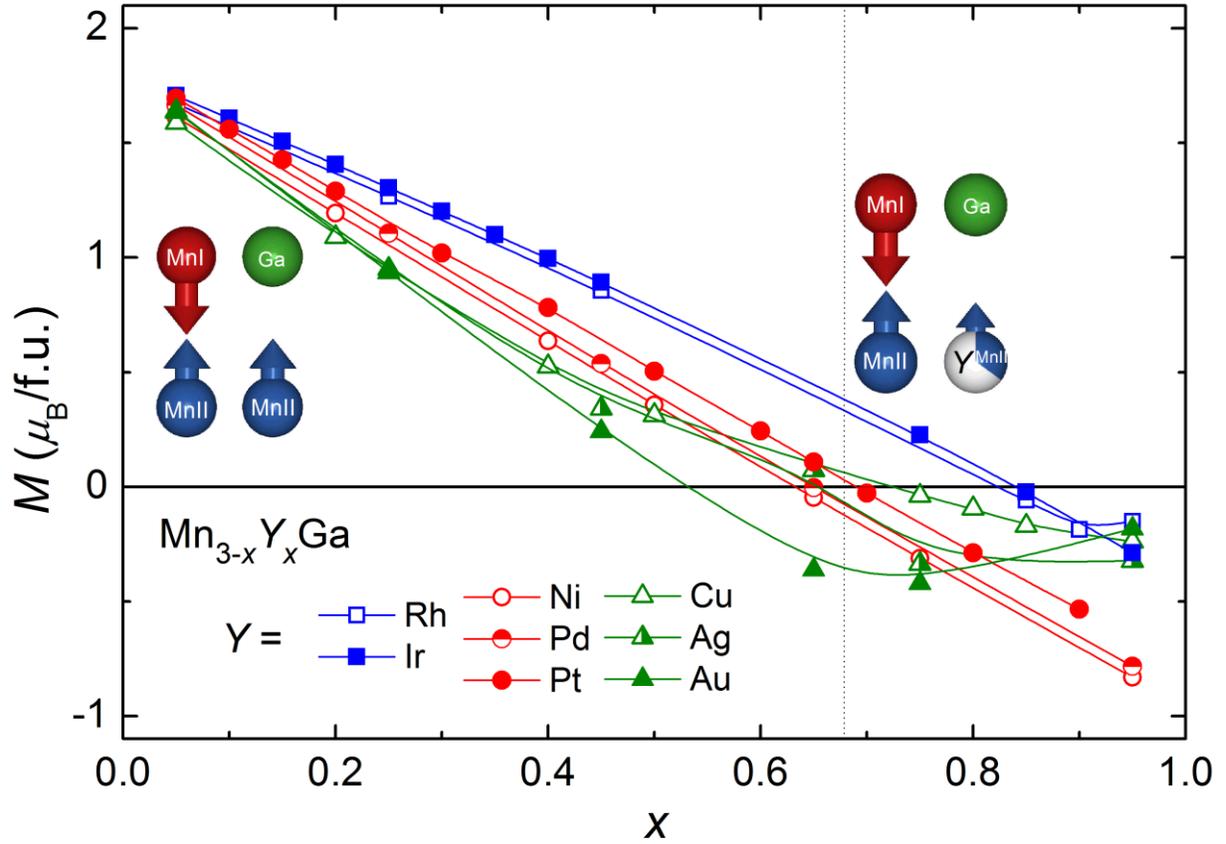

Figure 1. Compositional dependence of magnetization in $Mn_{3-x}Y_xGa$. $Y$ is a late transition metal, that is identified in the figure. Schematic diagrams of different magnetic states (shown by arrows) are shown in the respective compositional range. MnI (red spheres) and Ga (green spheres) sit on the sublattice-I whereas MnII (blue spheres) and $Y$ (gray spheres) sit on the sublattice-II. Antiferromagnetic exchange between the Mn atoms sitting on the different sublattices results in a zero magnetization when $Y$ partially substitutes MnII.

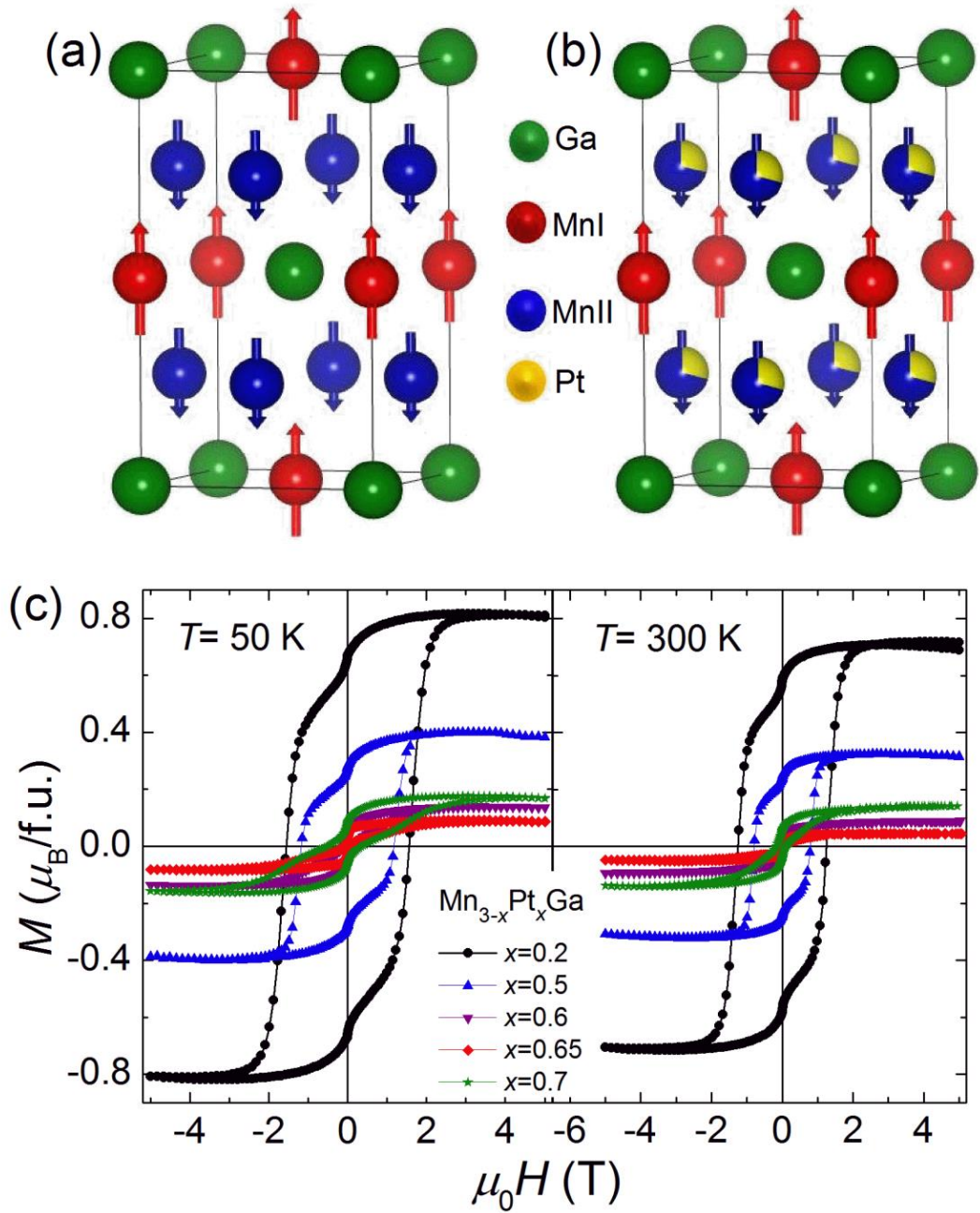

Figure 2. a) Crystal and magnetic structure of $Mn_3Ga$ with MnI (red spheres) sitting in the Mn-Ga plane and MnII (blue spheres) sitting in the Mn-Mn planes. The Ga atoms are shown as green spheres. The arrows represent the direction of the magnetic moment at each atom. b) Crystal and magnetic structure of $Mn_{2.4}Pt_{0.6}Ga$ with Pt atoms (golden spheres) partially replacing the MnII atoms. c) Field dependence of magnetization $M(H)$ for $Mn_{3-x}Pt_xGa$ measured at 50 K and 300 K.

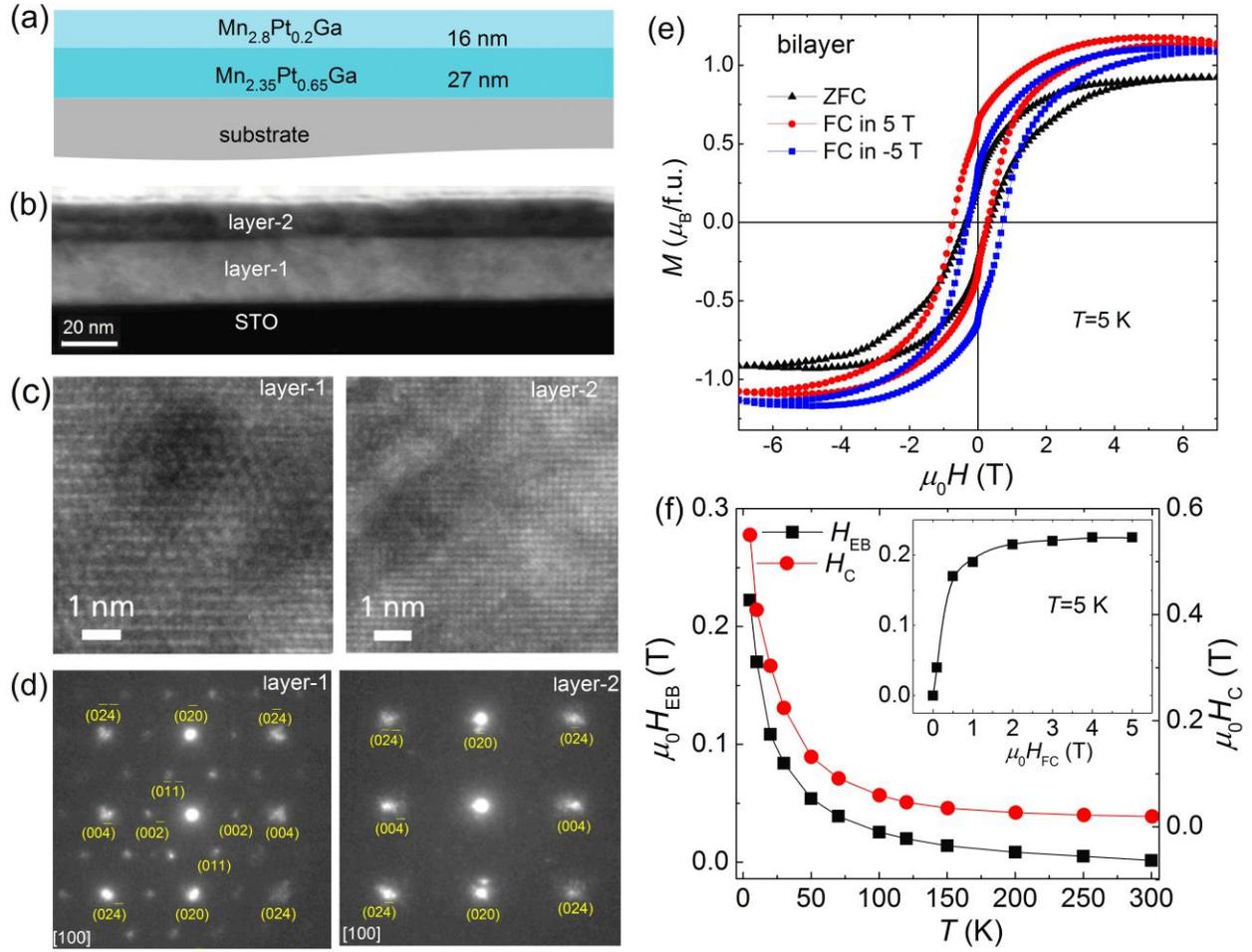

Figure 3. a) A schematic diagram showing bilayer of the present CFI Mn$_{2.35}$Pt$_{0.65}$Ga (27 nm) and FI Mn$_{2.8}$Pt$_{0.2}$Ga (16 nm). b) Cross-sectional scanning transmission electron microscope (STEM) image of the bilayer film. c) High resolution TEM image of layer-1 (left panel) and layer-2 (right panel). d) Selected area electron diffraction (SAED) patterns obtained in nano beam diffraction mode along the [100] zone axis of Mn$_{2.35}$Pt$_{0.65}$Ga (layer -1, left panel) and Mn$_{2.8}$Pt$_{0.2}$Ga (layer-2, right panel). e) $M(H)$ loops measured in zero field cooled (ZFC) and field cooled (FC) modes for the Mn$_{2.35}$Pt$_{0.65}$Ga/Mn$_{2.8}$Pt$_{0.2}$Ga bilayer film. f) Temperature dependence of coercive field ($H_C$) and exchange bias field ($H_{EB}$) for the bilayer film. The inset of **f** shows the cooling field dependence of EB.

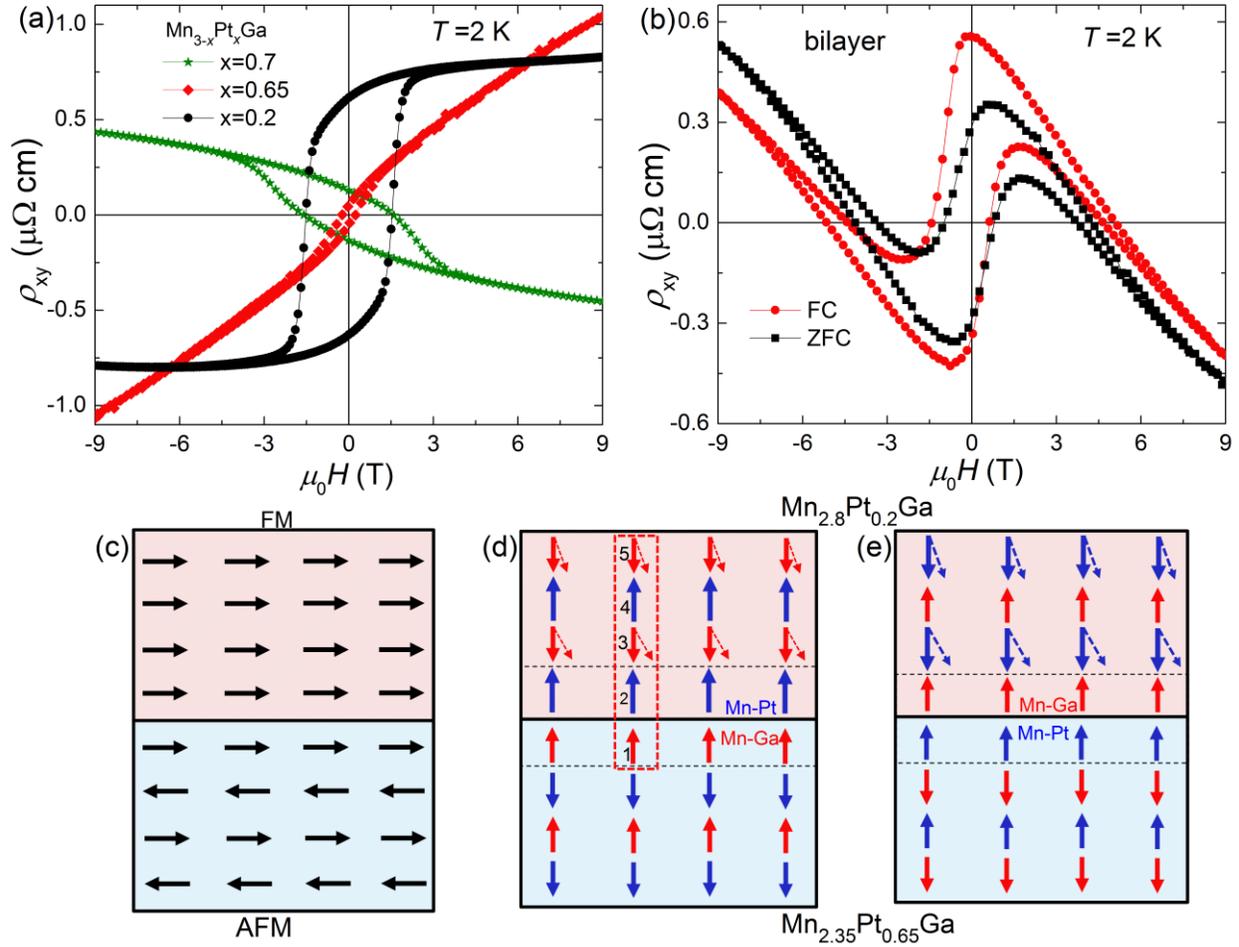

Figure 4. a) Anomalous Hall effect (AHE) versus perpendicular magnetic field for $Mn_{2.3}Pt_{0.7}Ga$, $Mn_{2.35}Pt_{0.65}Ga$ and $Mn_{2.8}Pt_{0.2}Ga$. All measurements are performed in the ZFC mode. b) AHE measured at 2 K in ZFC and FC modes for a $Mn_{2.35}Pt_{0.65}Ga/Mn_{2.8}Pt_{0.2}Ga$ bilayer sample. c) Schematic diagram of a conventional EB system with ferromagnetic and antiferromagnetic layers. Schematic diagram of the present CFI $Mn_{2.35}Pt_{0.65}Ga$ (bottom layer) and FI $Mn_{2.8}Pt_{0.2}Ga$ (top layer) with d) Mn-Ga/Mn-Pt and e) Mn-Pt/Mn-Ga interfaces.

# Supplementary Information:

# Compensated ferrimagne tetragonal thin films for antiferromagnetic spintronics

Roshnee Sahoo[1], Lukas Wollmann[1], Susanne Selle[2], Thomas Höche[2], Benedikt Ernst[1], Adel Kalache[1], Chandra Shekhar[1], Nitesh Kumar[1], Stanislav Chadov[1], Claudia Felser[1], Stuart S. P. Parkin[3] & Ajaya K. Nayak[1,3*]

[1]*Max Planck Institute for Chemical Physics of Solids, Nöthnitzer Str. 40, 01187 Dresden, Germany*
[2]*Fraunhofer Institute for Microstructure of Materials and Systems IMWS, Walter-Hülse-Str. 1, 06120 Halle, Germany*
[3]*Max Planck Institute of Microstructure Physics, Weinberg 2, 06120 Halle, Germany*

**Structural characterization:**

The out of plane (OP) XRD patterns for different Mn-Pt-Ga thin films are shown in Fig. S1a. All samples show (002) and (004) reflections of the tetragonal Heusler structure with space group $I4/mmm$, signifying an epitaxial growth of the films. This also confirms that a tetragonal crystal structure can be stabilized in Mn-Pt-Ga thin films. From the OP XRD measurements we found that with increasing Pt content the OP lattice parameter increases from $c = 7.14$ Å for $x = 0.2$ to $c = 7.37$ Å for $x = 0.7$ in $Mn_{3-x}Pt_xGa$. The increase in the OP lattice parameter with Pt content can be seen from the shifting of (002) and (004) reflections towards lower $2\theta$. The in-plane (IP) lattice parameter $a$ shows a small variation from 3.90 Å to 3.92 Å for the same compositional variation.

We have measured the low angle XRD or X-ray reflectivity (XRR) for the $Mn_{3-x}Pt_xGa$ thin films to determine the surface smoothness and thickness of the films, as shown in Fig. S1b. The observed oscillations of fringes are known as Kiesseg fringes, signifying high quality surface and good homogeneity of the present thin films. From the Kiesseg fringe spacing, the films thickness of $\approx 50$ nm is obtained after fitting the XRR curve. From the XRR fitting the average roughness is found to be less than 0.5 nm for all the films.

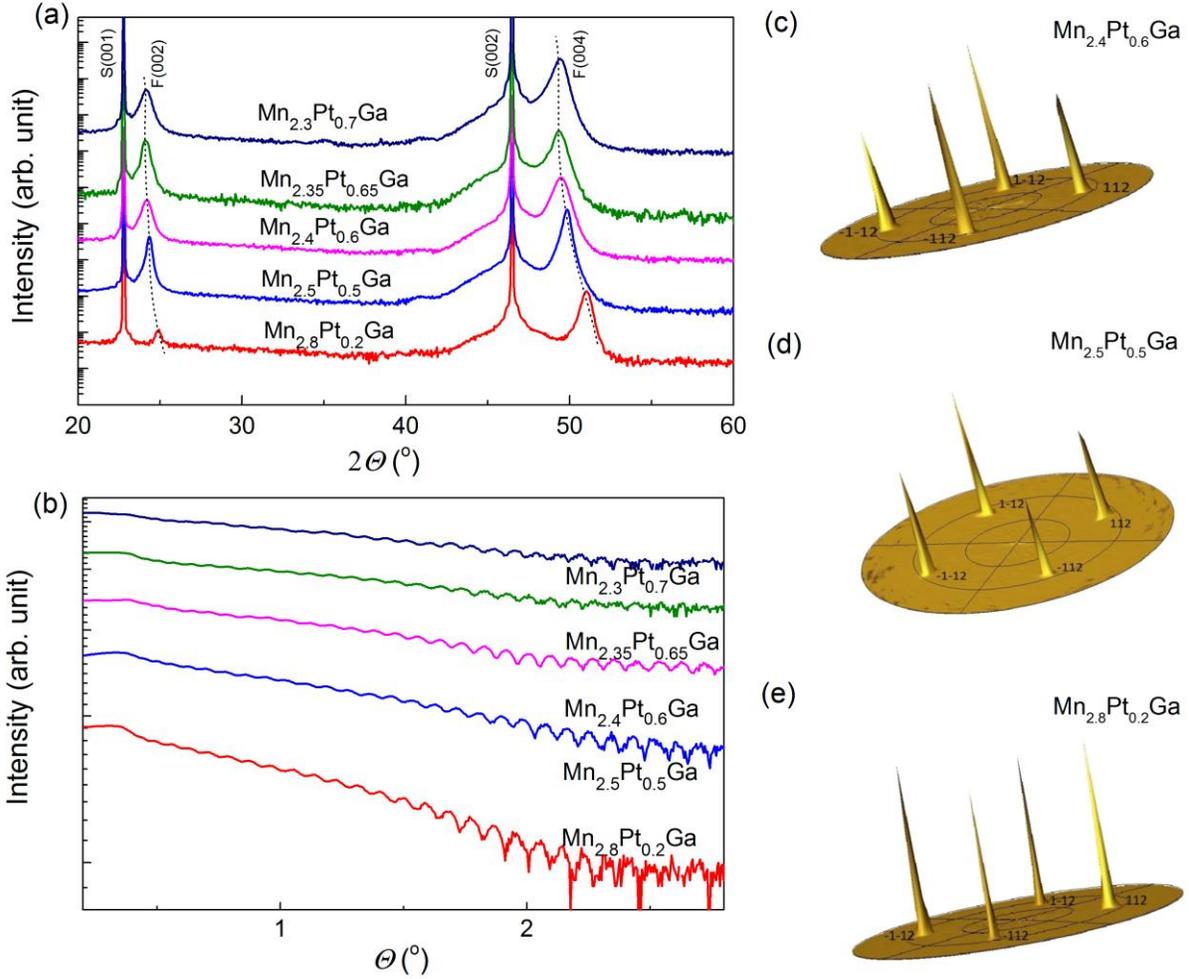

Figure S1. (a) Out of plane (OP) X-ray diffraction (XRD) patterns for different Mn-Pt-Ga thin films with thickness of 50 nm recorded at room temperature. S(001) and S(002) are the reflections from the SrTiO$_3$ (STO) substrate, whereas, F(002) and F(004) are reflections from the Mn-Pt-Ga films. (b) X-ray reflectivity diffractograms for different Mn$_{3-x}$Pt$_x$Ga thin films. Pole figures at (112) for (c) $x = 0.6$, (d) $x = 0.5$ and e, $x = 0.2$.

We have also performed in-plane XRD measurements to find out in-plane (IP) lattice parameter $a$ for different films. In particular, we concentrated on (112) peak for all the films. The peak is observed at 2θ=41.3°, 40.98° and 40.74° for $x = 0.2$, 0.5 and 0.6 respectively. The corresponding chi values are in between 52°-54°. The IP and OP lattice parameters for different films are given in Table S1. The IP lattice parameter varies from $a = 3.90(1)$ Å to a=3.93(1) Å by changing the composition from $x = 0.2$ to $x = 0.7$, respectively. This gives rise to a lattice mismatch of 0.1-0.6% with the substrate. The OP lattice parameter increases from 7.14 Å to 7.37 Å for the same compositional variation. It can also be noted that the intensity of the (002) peak oh these thin films increases with increasing Pt concentration, suggesting an enhanced atomic ordering for the films with higher Pt concentration.

The structural features were further analyzed by pole figures. XRD pole figure is a very useful tool to identify the presence of different phases and their exact orientations. In case of $DO_{22}$ structure, (112) peak possesses the maximum intensity. Therefore, pole figure measurements are performed for the (112) peak and are shown in Fig. S1c, d and e. The pole figure measurements show the diffracted intensity as a function of angle of rotation and tilt of the sample. For the present films four sharp peaks are observed at particular intervals. This confirms the four fold symmetry of the tetragonal crystal structure of the Mn-Pt-Ga thin films. This also establishes good peak distribution and crystalline orientation in these films. The four fold symmetry from pole figures explains good epitaxial nature in these films.

Table S1: Lattice parameters obtained from X-ray diffraction measurements in $Mn_{3-x}Pt_xGa$ films.

| $Mn_{3-x}Pt_xGa$ | $a$ (Å) | $c$ (Å) | (112) at $2\theta$ ° | $c/a$ |
|---|---|---|---|---|
| $x = 0.2$ | 3.90 | 7.17 | 41.30 | 1.84 |
| $x = 0.5$ | 3.90 | 7.30 | 40.98 | 1.87 |
| $x = 0.6$ | 3.92 | 7.35 | 40.73 | 1.87 |
| $x = 0.65$ | 3.93 | 7.38 | 40.64 | 1.87 |
| $x = 0.7$ | 3.92 | 7.37 | 40.84 | 1.88 |

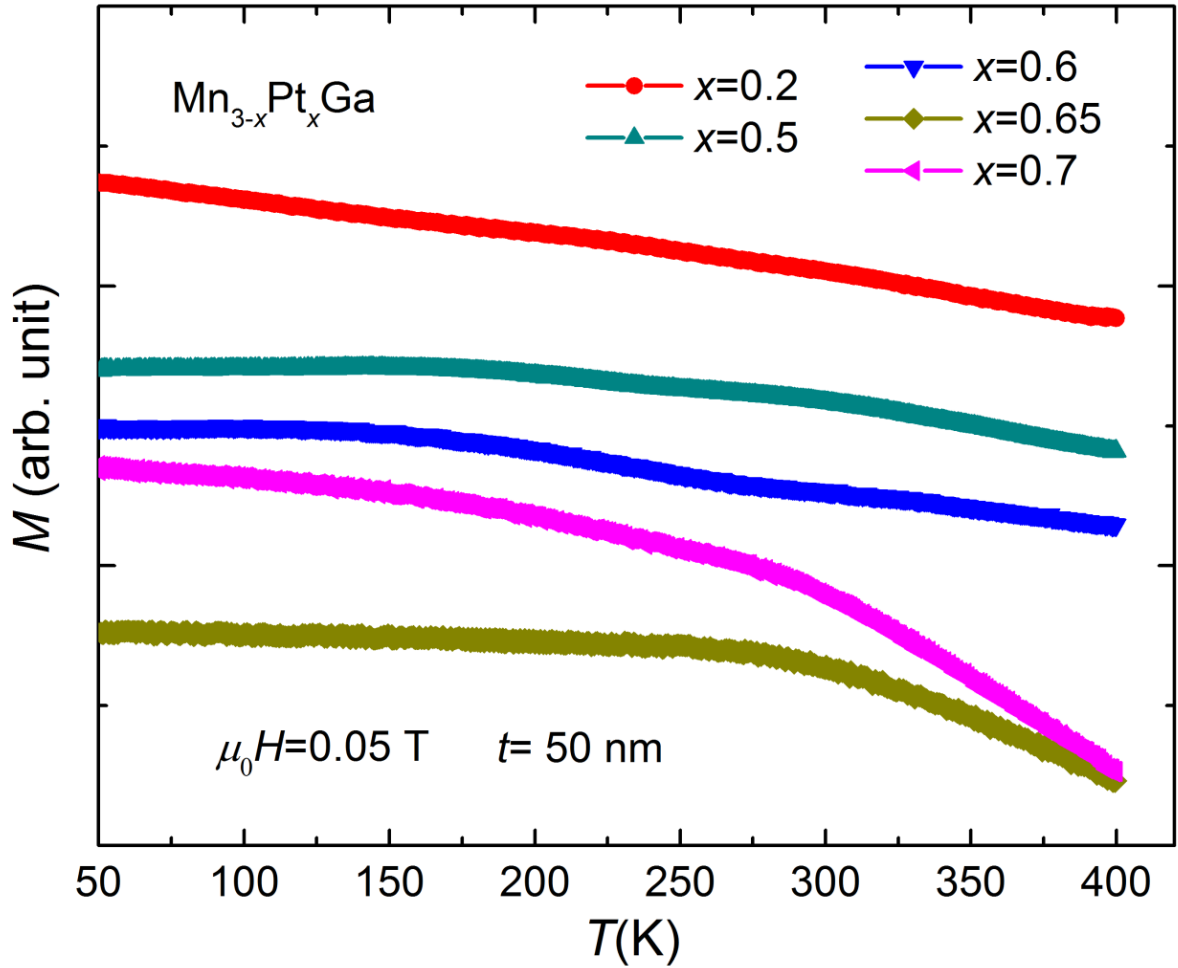

Figure S2. Temperature dependence of magnetization measured at 0.05 T for $Mn_{3-x}Pt_xGa$ ( $x = 0.2, 0.5, 0.6, 0.65$ and $0.7$) films.

Temperature dependence of magnetization ( $M(T)$ ) measurements performed in a temperature interval of 50-400 K in the OP direction of the present thin films with thickness 50 nm are depicted in Fig. S2. Films with lower Pt concentrations ( $x = 0.2, 0.5$ and $0.6$) exhibit a nearly temperature independent behavior up to 400 K, implying that the Curie temperature ($T_C$ of these films must lie at temperatures far above 400 K. For higher Pt substitution ( $x = 0.65$ and $0.7$) the transition shifts towards low temperature. Nevertheless, the transition is not complete up to 400 K. This implies that the $T_C$ is above 400 K.

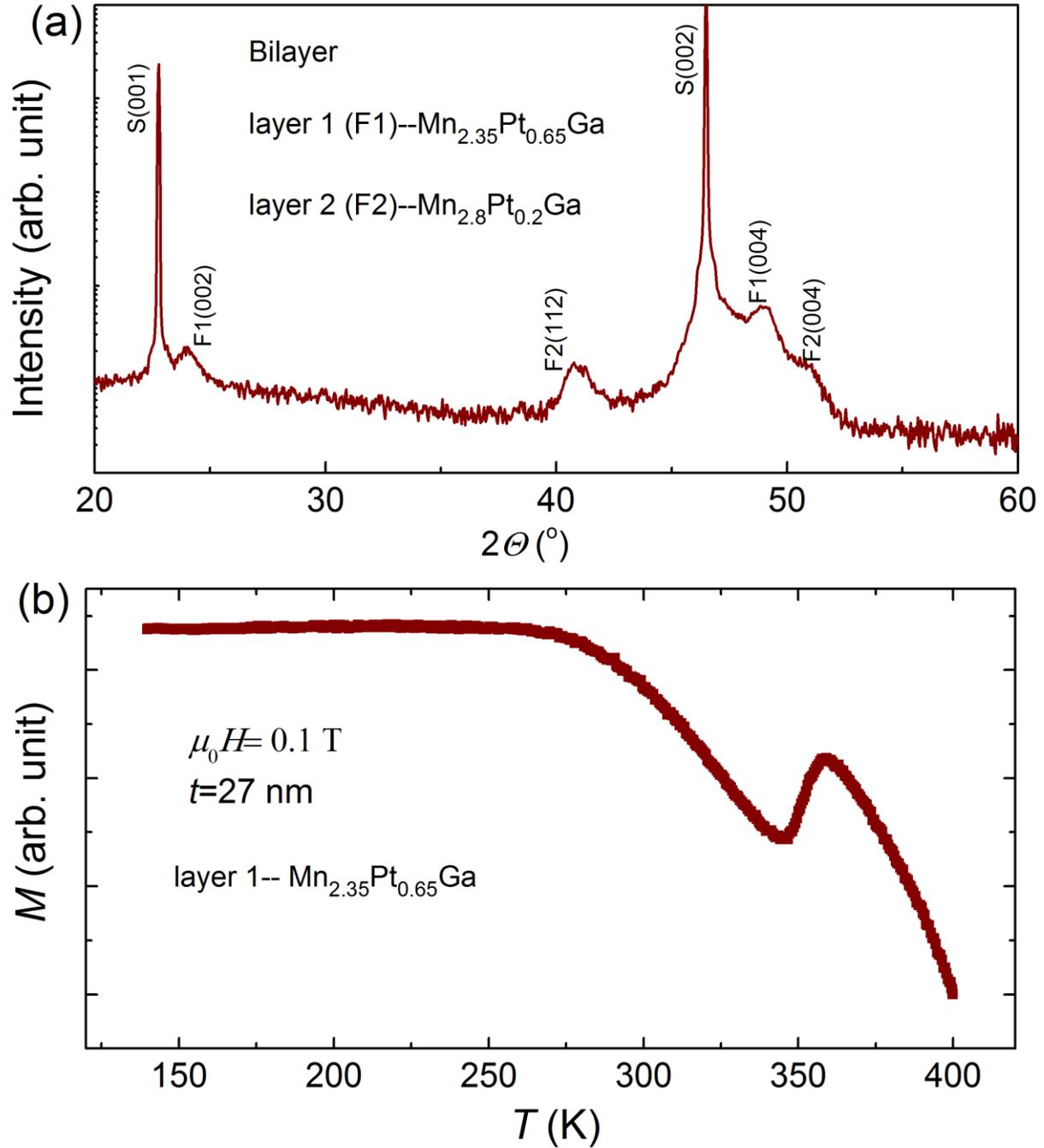

Figure S3: (a) XRD pattern for the $Mn_{2.35}Pt_{0.65}Ga/Mn_{2.8}Pt_{0.2}Ga$ bilayer film. (b) $M(T)$ curve for the 27 nm $Mn_{2.35}Pt_{0.65}Ga$ film.

The room temperature XRD pattern for the $Mn_{2.35}Pt_{0.65}Ga/Mn_{2.8}Pt_{0.2}Ga$ bilayer film is shown in Fig. S3a. As we can see, the XRD pattern clearly exhibits peaks corresponding to both the compensated ferrimagnetic film $Mn_{2.35}Pt_{0.65}Ga$ and ferrimagnetic film $Mn_{2.8}Pt_{0.2}Ga$. This confirms the formation of two distinguishable Mn-Pt-Ga layers. We also found that the ferrimagnetic layer (Mn$_{2.8}$Pt$_{0.2}$Ga) displays an extra (112) peak along with the OP (004) one, indicating existence of two different orientations.

Figure S3b shows $M(T)$ measured in a field of 0.1 T for the 27 nm thick Mn$_{2.35}$Pt$_{0.65}$Ga. We found that the $T_C$ of the compensated ferrimagnetic thin film can be reduced to $\approx$ 360 K by decreasing the thickness of the film from 50 nm to 27 nm. This helps us to study the exchange bias effect by field cooling the sample from 400 K.

**Theoretical calculations:**

The theoretical calculations are based on fully-relativistic density functoinal theory as implemented in the Munich SPR-KKR package, wherein chemical disorder is described by means of the coherent potential approximation.[1] The experimental lattice parameters of Mn$_3$Ga constitute the structural model of the theoretical computations in which one Mn atom in the Mn-Mn plane (Wyckoff position 8c, 4d/4c) is substituted by a late transition metal $Y$ = Ni, Cu, Rh, Pd, Ag, Ir, Pt, Au. The theoretical calculations motivated the experimental work on Mn-Pt-Ga, that then was used as structural model for further calculations, assuming atomic occupations as Ga on 4a, Mn on 4b, Mn/Pt on site 4c and Mn on site 4d. From the above study we have established that a compensated magnetic state can be achieved in Mn-Pt-Ga based tetragonal thin films.

Tetragonal Heusler alloys have not found to be purely half-metallic showing a real gap in one spin-channel, yet. Nevertheless, spin-polarized currents may in theory be obtained following the scheme of *spin-selective electron localization*, as proposed by Chadov *et al.*.[2] A general discrepancy between spin-polarizations as obtained from the density of states (DOS),

$$P_{\alpha\alpha} = \frac{\sigma_{\alpha\alpha}^{\uparrow} - \sigma_{\alpha\alpha}^{\downarrow}}{\sigma_{\alpha\alpha}^{\uparrow} + \sigma_{\alpha\alpha}^{\downarrow}}, \qquad (1)$$

and from conductivity calculations,

$$P_{n_{\text{F}}} = \frac{n_{\text{F}}^{\uparrow} - n_{\text{F}}^{\downarrow}}{n_{\text{F}}^{\uparrow} + n_{\text{F}}^{\downarrow}}. \qquad (2)$$

Therefore, we additionaly evaluate the spin-polarization defined in terms of the spin-projected diagonal components of DC residual conductivity tensor, $\sigma_{\alpha\alpha}^{\uparrow(\downarrow)}$, where $\alpha$ is the spatial component – $x$, $y$ or $z$. The underlying computation is based on the Kubo-Greenwood formalism [3, 4] and implies the relativistic spin-projection technique. [1, 5]

As it turns out, the spin-polarizations $P_{\alpha\alpha}$, defined in the sense of Eq. 1, exhibits not only a different amplitude, but also an opposite sign, as compared to $P_{n_{\text{F}}}$ defined in the sense of Eq. 2 (see Fig. S4a). This is caused by the fact, that the mobility of the conducting electrons in one of the spin-channels (marked as blue on the corresponding DOS and Bloch spectral function (BSF) shown in Fig. S4b is much lower than in another one (marked as red), despite that the overall quantity of these electrons is obviously larger.

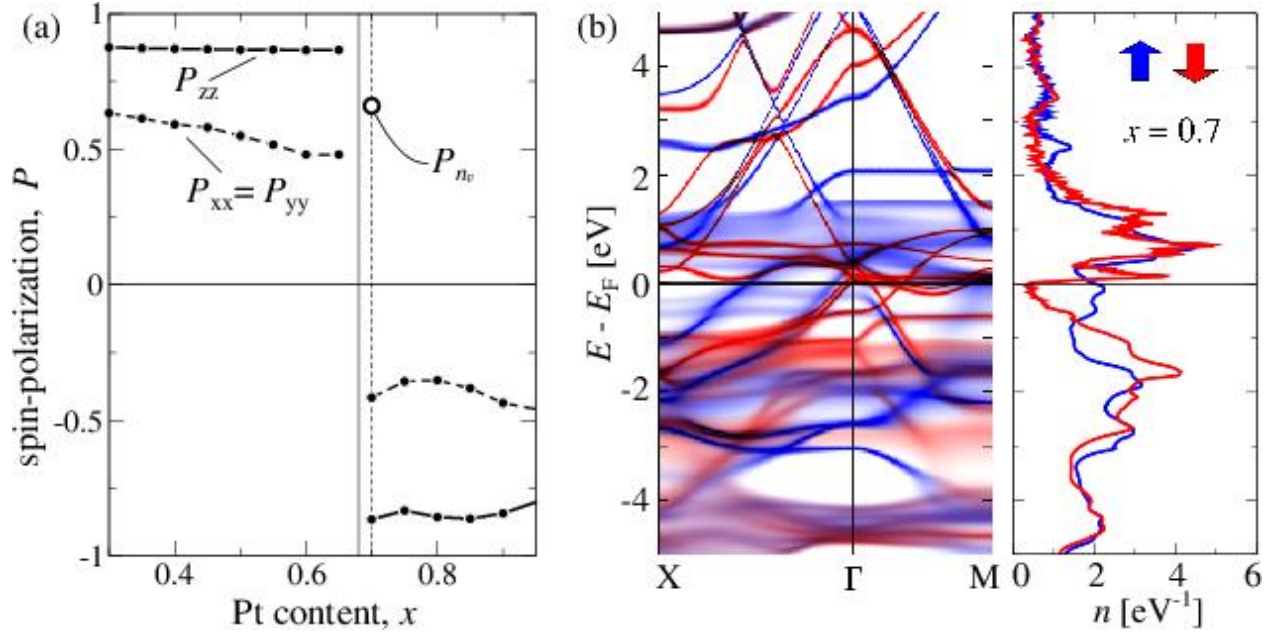

Figure S4: **Spin-polarizations. a,** Spin-polarizations calculated as a function of Pt content $x$ in $Mn_{3-x}Pt_xGa$. $P_{xx} = P_{yy}$ and $P_{zz}$ are the spatial components of spin-polarization defined in Eq. 1. $P_{n_F}$ at $x = 0.7$ is defined using Eq. 2. The dashed vertical line at $x = 0.7$ corresponds to the calculated composition, which appears to be the closest to the experimental compensation point. **b,** Spin-projected BSF (red and blue indicate spin-down and spin-up states, respectively) along the X-$\Gamma$-M path and the spin-projected total DOS, $n$, computed for $x = 0.7$.

It can be understood from the detailed BSF plot for $x = 0.7$ in $Mn_{3-x}Pt_xGa$, that in the vicinity of $E_F$ the states of the spin-up channel (blue) are substantially more broadened compared to another spin channel (red). This broadening stems from the chemical Mn/Pt disorder in the Mn-Pt plane. This interesting effect, namely the "spin-selective localization" of conducting electrons was theoretically proposed for $Mn_{3-x}Y_xGa$ series, where Y is a late $3d$ transition metal element.[2]

As it is expected, the tetragonally-distorted unit cell leads to anisotropic spin-polarization: $P_{xx} = P_{yy} \neq P_{zz}$. At the same time, the largest and relatively constant (as a function of $x$) value has the $P_{zz}$ component ($:\pm 0.87$), which corresponds to the electric current along the tetragonal $c$ axis. Similar results were obtained in our previous studies on similar systems.[2,6] By going through the compensation point at $x \approx 0.68$, all components of spin-polarization change the sign, but not the amplitude, which means that the compensation point composition has the same high spin-polarization along the $c$ axis.